# Terahertz source-on-a-chip with decade-long stability using layered superconductor elliptical microcavities

Mingqi Zhang[1], Shungo Nakagawa[2], Yuki Enomoto[2], Yoshihiko Kuzumi[2], Ryuta Kikuchi[2], Yuki Yamauchi[2], Toshiaki Hattori[2], Richard A. Klemm[3], Kazuo Kadowaki[2], Takanari Kashiwagi[2], and Kaveh Delfanazari[1,*]

[1] *James Watt School of Engineering, University of Glasgow, Glasgow G12 8QQ, UK*
[2] *Institute of Material Science, University of Tsukuba, 1-1-1 Tenoudai, Tsukuba, Ibaraki 305-8573, Japan*
[3] *Department of Physics, University of Central Florida, 4111 Libra Drive, Orlando, Florida 32816-2385, USA*
\* *Correspondence*:   Kaveh.Delfanazari@glasgow.ac.uk      Dated: 28062025

**Abstract. Coherent, continuous-wave, and electrically tunable chip-scale terahertz (THz) sources are critical for emerging applications in sensing, imaging, spectroscopy, communication, space and quantum technologies. Here, we demonstrate a robust source-on-a-chip THz emitter based on a layered high-temperature superconductor, engineered with an elliptical microcavity and capable of sustained coherent emission over an unprecedented operational lifetime exceeding 11 years. This compact THz source operates up to ~60 K (with $T_c \approx 90$ K), delivering stable radiation in the 0.7–0.8 THz range, with on-chip electrical tunability from 100 GHz to 1 THz. Coherence arises from the phase-locked oscillation of intrinsic Josephson junctions (IJJs) arrays, resonantly coupled to transverse electromagnetic modes within the cavity, analogous to a laser cavity, yielding collective macroscopic oscillations. THz emission remains detectable across a ~0.5 m free-space open-air link at room temperature. We analyse the cavity-mode structure and extract THz photon generation rates up to ~503 photons fs$^{-1}$ in cryogenic conditions and 50–260 photons ps$^{-1}$ over-the-air. These results establish long-term coherent THz emission from superconductors and chart a viable path toward scalable, tunable, solid-state coherent THz laser-on-a-chip platforms, especially for future classical and quantum systems.**

**Introduction** Terahertz (THz) waves, spanning 0.1–10 THz between microwave and optical frequencies, have gained significant attention in recent decades due to their potential in fundamental research and diverse applications. Beyond classical uses in biomedical imaging, chemical identification, nondestructive inspection, and next-generation communication, recent advances in quantum technology suggest the feasibility of THz quantum communication and quantum sensing [1-10]. The advantages of THz waves, including high stability in dust, fog, and atmospheric turbulence compared to free-space optical links, and higher operating temperatures in quantum computing compared to microwaves, highlight the untapped potential of the THz region as a key research focus [11]. However, efficient THz hardware, including sources, modulators, and detectors essential for practical applications, is still in its early stages [12-18]. High-temperature superconducting $Bi_2Sr_2CaCu_2O_{8+\delta}$ (BSCCO) emitters, composed of two-dimensional van der Waals (vdWs) layers with naturally formed intrinsic Josephson junctions (IJJs) arising from alternating $Bi_2O_2$ insulating and $CuO_2$ superconducting layers, have been shown to generate broadband terahertz (THz) radiation spanning from 150 GHz to 11 THz, using both on-chip and off-chip detection techniques. The mesa structures in these devices support a macroscopic coherent electromagnetic state through cavity resonance, wherein a large number of stacked IJJs are phase-synchronized, resulting in coherent THz emission, a phenomenon extensively reported and validated in foundational studies [19–22]. For sustained over-the-air applications, the stability and longevity of a THz source are as crucial as its power and bandwidth for practical viability. This study introduces several key novel contributions to solid-state THz technology, including: (i) robust, long-lasting, and electrically tunable elliptical cavity BSCCO THz source, emitting within the 100 GHz–1 THz range, with intense radiation at 700–800 GHz. Fabricated and measured in early 2012, and remeasured in late 2023, the device retains its superconducting phase transition, electrical response, and THz output power, demonstrating exceptional long-term stability and operational resilience. (ii) The THz emissions were detected across a 44 cm free-space channel nearly 12 years after fabrication, highlighting the device's viability for sustained cryogenic and free-space applications. (iii) The unique elliptical cavity design optimises transverse electromagnetic modes, enabling efficient photon generation and tunable THz emission from 100 GHz to 1 THz. (iv) High THz photon generation rate and achieving ~503 photons per femtosecond ($fs^{-1}$) in cryogenic conditions and 50–260 photons per picosecond ($ps^{-1}$) over-the-air transmission, highlighting the device's importance for various applications, especially in classical or quantum communication and computation.

**Coherent THz source fabrication and measurement setup** The BSCCO elliptical cavity THz source was fabricated from a BSCCO single crystal grown using the travelling solvent floating zone (TSFZ) method [23, 24]. The crystal was exfoliated into thin flakes using Scotch tape, and a flat, optimally thick flake was selected, transferred, and affixed to a sapphire substrate. A focused ion beam (FIB) system was employed to etch the elliptical geometry, creating three elliptical cavity structures, named devices 1, 2, and 3, arranged in a parallel array on the surface of the BSCCO chip, as shown in Fig. 1(a). The three elliptical cavities are nearly identical in size, with dimensions $a≈245$ μm, $b≈52$ μm, and $d≈1$ μm, as shown in Fig. 1(b). The cavity (mesa) thickness $d$ was estimated from the number of Josephson junctions using the AC Josephson effect, with each junction $d_j≈1.533$ nm [25], as the FIB may not etch through the entire BSCCO flake on the substrate. Gold wires were attached to the top of each elliptical device and the edge of the BSCCO crystal to apply an electric field across the IJJs. The fabrication process is detailed in Appendix Fig. A1(a), and a scanning electron microscope (SEM) image of the device is shown in Fig. A1(b). The sample was mounted to the cold finger using silver paste in 2012, copper clamps and thermal grease in 2023, then cooled in a $^4$He flow cryostat.

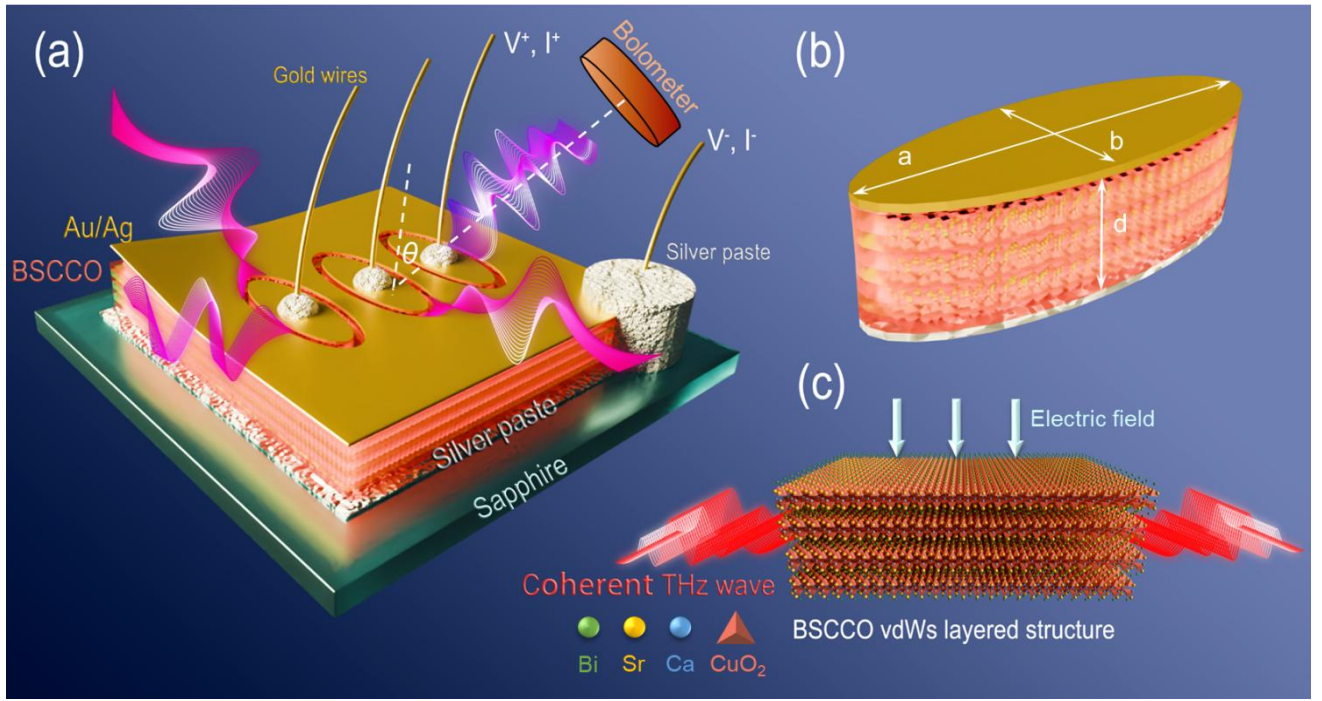

**Figure 1** (a) Schematic of the BSCCO elliptical cavity THz source array. (b) Detailed schematic of an individual elliptical cavity on the device, with dimensions $a≈245$ μm, $b≈52$ μm, and $d≈1$ μm. (c) Illustration of the BSCCO van der Waals (vdWs) layered structure, forming nanoscale intrinsic Josephson junctions within the BSCCO cavity. In this study, the devices were operated and measured individually within the array structure.

For basic measurements such as temperature-dependent resistance and current-voltage characteristics, a voltage source was used to apply an external DC voltage across the device. The bias voltage and current were monitored using two separate digital multimeters. Radiation output was detected using a Si-bolometer (for the 2012 measurement) or an InSb hot electron bolometer (HEB) (for the 2023 measurement), with an optical chopper operating at 70 Hz for amplitude modulation, employed in the lock-in technique. This is a standard method for THz measurement widely used for the characterization of BSCCO devices [18-22]. The setup was positioned in front of the THz window of $^4$He flow cryostat, as illustrated in Fig. A2 in Appendix A. The angle between the normal of the device and the bolometer is defined as $\theta$, illustrated in Fig. 1(a). The frequency measurement setup and results are discussed in detail in Section IV.

**Electromagnetic wave emission and coherent THz source sustained over 11 years** For the first time, we have successfully measured and confirmed stronger THz radiation output from a device almost 12 years after its fabrication, along with its experimentally determined frequency. In this section, we will discuss the measurement results of critical temperature, current-voltage characteristics (IVCs), and emission from the 2012 and 2023 measurements. Comparing the IVCs of device 1 in Fig. 2(a) at different bath temperatures for the 2012 (red) and 2023 (blue) measurements, both sets exhibit the typical hysteresis and branching structure characteristic of IJJs stacks [25]. The sharp voltage jump from <0.7 V to >0.9 V, $I \gtrsim 40$ mA, observed in the 2012 IVCs reflects the generation of hot spots, leading to a significant temperature increase in the device [26]. On the return branches, the back-

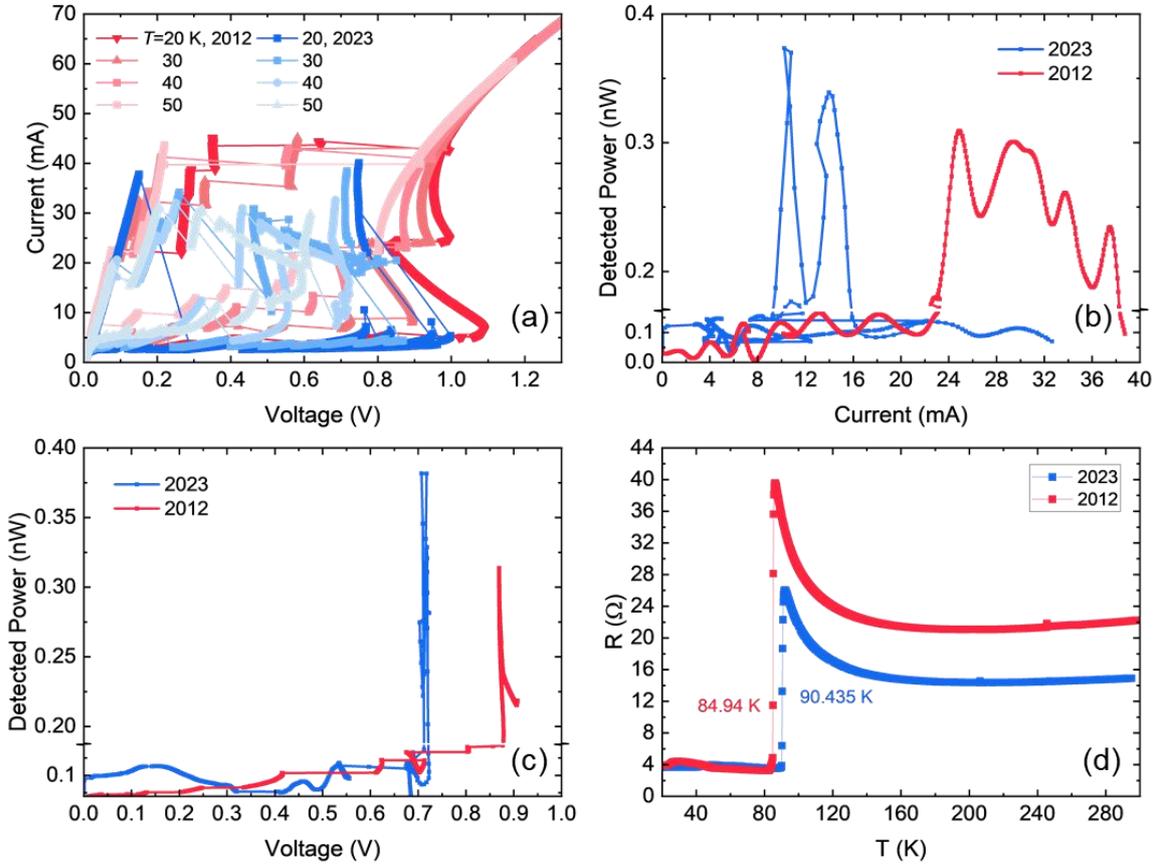

**Figure 2** (a) Current-voltage characteristic for device 1 at different bath temperatures in 2012 (red set) and 2023 (blue set). (b) Over-the-air detected power versus bias current for device 1 at 40 K, at $\theta=0°$ in 2023 and $\theta=30°$ in 2012. (c) Over-the-air detected power versus bias voltage for device 1 at 40 K, at $\theta=0°$ in 2023 and $\theta=30°$ in 2012. (d) Temperature-dependent resistance (RT) curves for device 1 in 2012 and 2023.

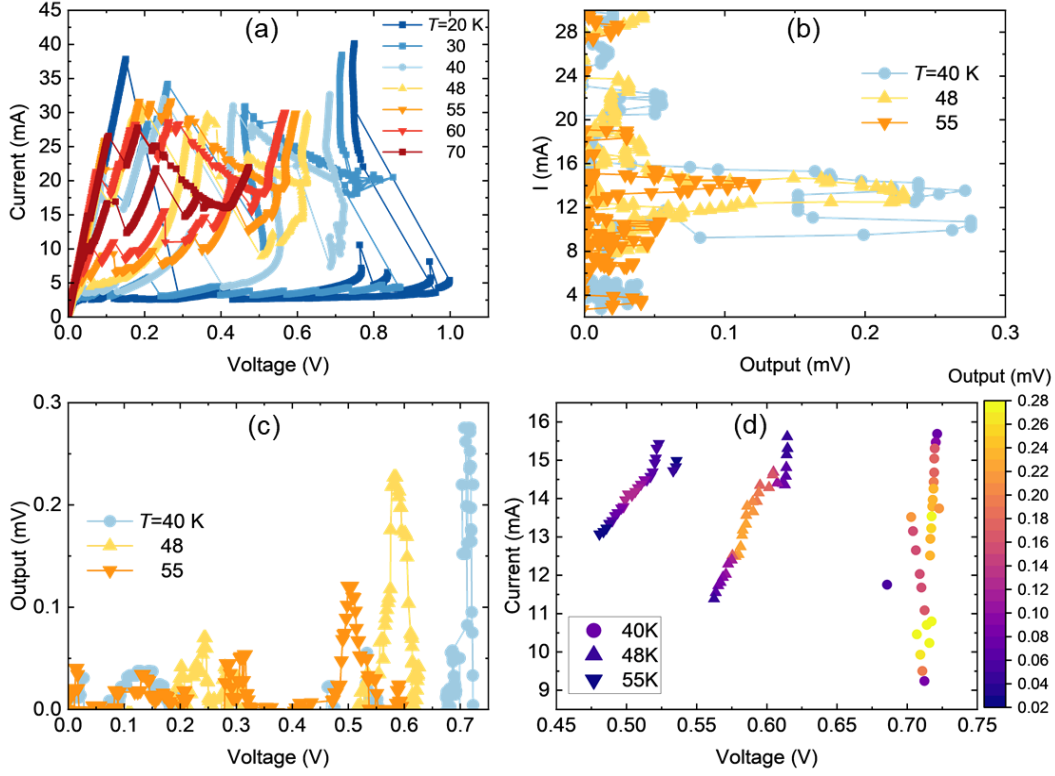

**Figure 3** (a) Current-voltage characteristics (IVCs) at different bath temperatures for device 1 in 2023. Output from device 1 detected by the HEB versus (b) bias current at different temperatures, at $\theta=0°$ in 2023, and vs. (c) bias voltage at different temperatures, at $\theta=0°$ in 2023. A standard median filter has been used to reduce the background noise level in these figures. (d) IVCs at the THz radiation range at different temperatures.

bending in the 2023 measurements is less pronounced than in 2012, across all bath temperatures. Considering the sudden voltage jump to its highest value at bath temperatures of 20 K and 30 K, the thermal release of the device's local temperature in 2023 appears to be improved compared to 2012 [26, 27]. Therefore, using copper clamps to securely attach the device directly to the cold finger with thermal grease beneath the sapphire substrate enhances heat removal during the measurement, compared to using silver paste. The emission directed to the Si or HE bolometer can be observed through the bolometer's output voltage, $V_{output}$. However, the optical responsivity $\alpha$ differs between the two types of bolometers. To compare the radiation output in 2012 and 2023, we estimated the output voltage in terms of the radiation power reaching the detector at a bath temperature $T_b$=40 K, as shown in Fig. 2(b) and (c). The relationship between the output voltage and detected power is given by:

$$P = 2\sqrt{2}\frac{V_{output}}{\alpha} \quad (1)$$

where the optical responsivity $\alpha_{HEB}$=3.3 mV/nW for the 2023 HEB detection and $\alpha_{Si}$=11 mV/nW for the 2012 Si-bolometer detection, both calibrated by black body radiation, and with coefficient number $2\sqrt{2}$ [28]. The emission occurred at a lower bias voltage and current in 2023, which is attributed to the doping change of the BSCCO crystal, consistent with the critical temperature increase from ~85 K in 2012 to ~90 K in Fig.2 (d). This suggests the device was nearly optimally doped in 2023 [29]. Considering the emission in 2023 is more substantial than in 2012 at 40 K, the device achieved higher radiation efficiency in 2023, with lower bias power and higher detected power. Moreover, the operating temperature of this device in 2023 reached $T_b$=55 K, still achieving about half of the highest output at $T_b$=40 K, as shown in Fig. 3(b) and (c). While the bias current remains relatively stable across different bath temperatures in Fig. 3(b), the emission occurred at a lower bias voltage along the back-bending region at higher bath temperatures, as seen in Fig. 3(c). The back-bending in the 2023 IVCs in Fig. 3(a) became increasingly less noticeable at higher temperatures and eventually disappeared at around 60 K. This behaviour is related to Newton's law of heating, which leads to the device overheating at lower bias voltages as the bath temperature increases [30]. The RT curves and IVCs for the other devices from the 2012 measurement are provided in Fig. A3. It is worth noting that the development of BSCCO-based THz sources has undergone intensive theoretical and experimental investigation over the past two decades, ranging from fundamental studies of IJJ dynamics to practical device geometries and radiation enhancement schemes, and more recently to application-oriented integration efforts [31–48]. From Fig.3 (d), we can gain a clearer view of the radiation intensity at different bias points at different temperatures. The double radiation peaks in Fig.3 (b) at 40 K are related to the jump into a different branch at the back-bending region caused by the unreleased hot spot, as illustrated in Fig.3 (d). Together with the radiation region I-V with less pronounced back bending at higher temperature, these characteristics match the former study, which suggests that the higher radiation intensity at 40 K is related to the presence of a hot spot [45]. Due to long-term storage, the top electrode gold wires, attached with silver paste on devices 2 and 3, were found to be broken in 2023. Therefore, improving the electrical feeding method and device packaging is critical for future device generations. This paper focuses specifically on THz emitters with elliptical cavities, investigating the mechanisms of electromagnetic wave radiation and the associated cavity modes. However, we have also observed THz emission from BSCCO devices fabricated concurrently on a separate chip. A detailed analysis of those results will be presented in a forthcoming publication.

**Free-space THz radiation frequency measurements over-the-air and analysis** Considering the free space detected power of device 1 after a 10 cm over-the-air link via the HEB setup in Fig. A2 (b), which is around 0.5 nW and highly sensitive to changes in bias voltage, this presents a significant challenge for frequency measurement over long-distance over-the-air channels. To address this, we replaced the DC source (Keithley 2400 Sourcemeter) and optical chopper with a wave generator (Agilent 33220A), which superimposes a 1 kHz sinusoidal wave onto the DC voltage with a 70 mV$_{pp}$ peak-to-peak amplitude, as shown in Appendix A, Fig. A4. This setup provides an accurate bias voltage and enables the fast switching of the signal pulse simultaneously. The frequency was measured using a Michelson interferometer in over-the-air (room temperature), as shown in Fig. A4. Notably, the THz wave measured from our device passed through a free-space link of over 44 cm, and only half of the power radiation reached the detector through the beam splitter, which is promising for applications requiring wireless channels over-the-air.

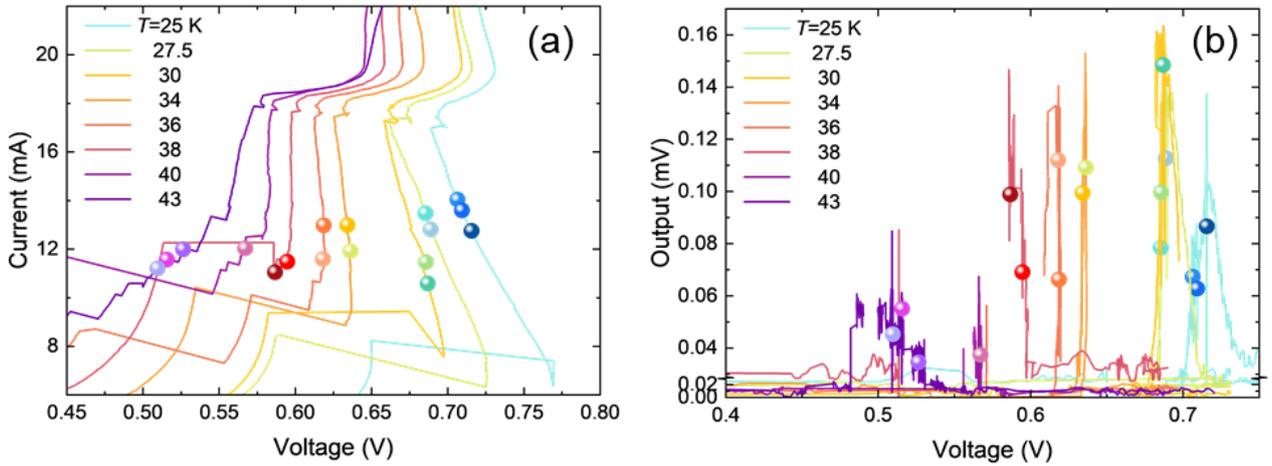

**Figure 4** (a) Current-voltage characteristic for device 1 at different bath temperatures in 2023. The highlighted points are the bias points for frequency measurement at $\theta=0°$. (b) The output voltage from the HEB detector at different temperatures, and $\theta=0°$. The highlight points are output points for frequency measurement.

The IVC curves of device 1, driven by the wave generator at different bath temperatures, are shown in Fig. 4(a). The highlighted points on each curve indicate the bias points corresponding to the frequency spectrum measurement, as shown in Fig. 5 (a). These bias points were selected around the output measurement peaks from the HEB, illustrated in Fig. 4(b). In this measurement setup, the lowest temperature at which THz radiation was measured was 25 K, with strong radiation detectable between 30 K and 40 K. The highest detected power was approximately 0.137 nW (~0.16 mV output voltage) after the 44 cm propagation at $\theta=0°$ with the setup in Fig. A4. Similar measurements were also conducted at $\theta=30°$, as shown in Fig.5 (b) with the IVCs and output in Appendix A, Fig. A5. From the frequency spectrum in Fig.5, THz radiation was observed between 100 GHz and 1 THz, with the most powerful radiation frequency tunable around 650-800 GHz. The sampling frequency of the Michelson interferometer limited the resolution of the spectrum. In Fig. 5(a) and Fig. 5(b), we observe a slight shift in the frequency peak, influenced by both bias voltage and bath temperature. The relationship between frequency and bias voltage at the same bath temperature is consistent with the ac-Josephson relation, given by

$$f = \frac{2eV}{hN} \qquad (2)$$

where $e$ is the electronic charge, $V$ is the voltage across the junctions, $h$ is Planck's constant, and $N$ is the number of junctions. The number of active junctions in device 1 changes between 300 and 500, and is inversely related to the bath temperature, as shown in Fig. A6. The tunable relation between $f$ and $N$ at each temperature in Fig. A6 is similar to the branch effect observed in previous studies [31].

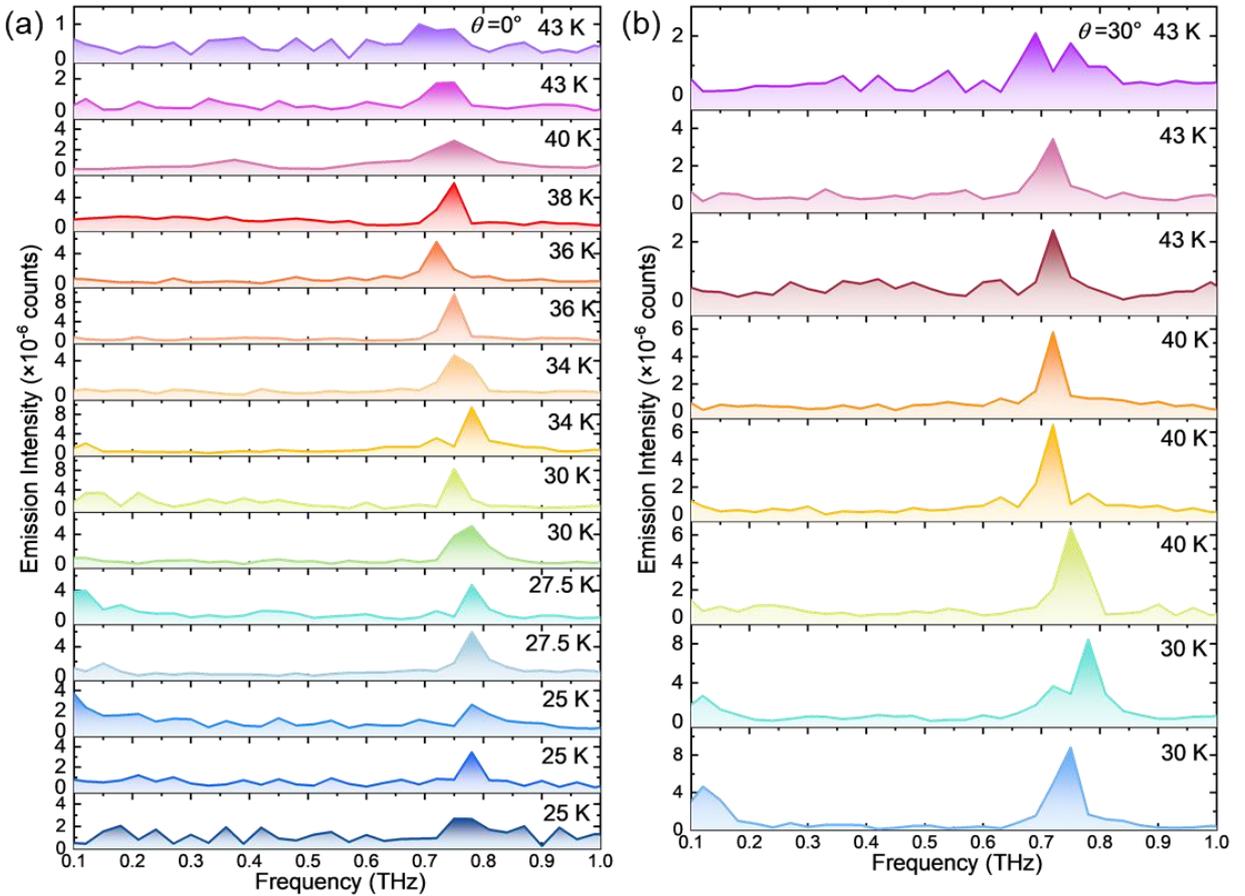

**Figure 5** (a) Frequency spectra from 100 GHz to 1 THz for the elliptical cavity at different bath temperatures at $\theta=0°$. The intensities are scaled by factors of 3, 3, and 2 for 43 K and 40 K. Grey dashed lines indicate the frequency shift trend. (b) Frequency spectra from 100 GHz to 1 THz for the elliptical cavity at different bath temperatures at $\theta=30°$. The intensities are scaled by factors of 3, 2, and 3 for 43 K. The intensity in (a) and (b) is offset for clarity.

The variation in the number of active junctions in different measurements at the same temperature may be due to jumps into different branches in the back-bending region. In addition to the strong frequency peaks from the ac-Josephson effect, smaller peaks around 120 GHz and 210 GHz are observed in the measurements shown in Fig. 5 (for a zoom-in area see Fig. A9). Weak peaks in the 0.3–0.6 THz range also appear at certain temperatures. While the geometry of the BSCCO mesa can activate cavity modes alongside the ac-Josephson effect [22], further investigation is needed to fully determine the origin of the low-intensity peaks in the spectrum. Studies using newly fabricated elliptical cavity devices with higher-intensity radiation will be crucial for a deeper understanding of these peaks. By rotating the sample at different angles $\theta$, as defined in Fig.6 (b), we plot the angle-dependent radiation pattern of device 1 at $T_b$=45 K as shown in Fig.6 (a) (red dots). The angular range in this figure was constrained by the hardware's degrees of freedom. The power is calculated from the highest output voltage at each angle of the HEB by Eq.1. Since the primary radiation intensity is concentrated around 700-800 GHz, the resonance frequency 2D far-field pattern of the elliptical cavity structure was simulated in COMSOL within the same frequency range to analyse the electric field distribution. The 2D far-field pattern, taken from the same direction as the measurement, shows a square relationship with power. To identify the optimal radiation mode and frequency for comparison with the radiation pattern, we employed both mathematical analysis and simulations in the following section. Building on previous research, a BSCCO THz emitter can be modelled as an antenna with Neumann boundary conditions along the edge of the ellipse, where the normal derivative of the wave function vanishes at the boundary [32].

In the simulation model, the boundary can also be defined using the surface current density boundary condition [33]. The antenna's TM mode can be estimated from the modified Mathieu function, applying the Neumann boundary condition for the TM(m, r) mode as given by [49]

$$0 = \begin{cases} \partial_\mu Ce\left(\mu_0, q_{mr}^{(1)}\right) \\ \partial_\mu Se\left(\mu_0, q_{mr}^{(2)}\right) \end{cases} \quad (3)$$

where $Ce$ and $Se$ represent the even and odd solutions, $\mu_0 = \text{atanh}\left(\frac{b}{a}\right)$, is the eccentricity of the ellipse. Note that the TM mode described here differs from the 2D model in [49], that is based on the microstrip antenna model and the boundary conditions outlined in [32]. The resonance frequency $\omega$ is related to [49]

$$q_{mr} = \frac{k^2 l_f^2}{4} \tag{4}$$

$$k^2 = \frac{n^2 \omega^2}{c^2} \tag{5}$$

with the focal length of the ellipse $l_f$, the refractive index $n$ of BSCCO, and the speed of light $c$. The relative permittivity of BSCCO is set to be $n^2$=17.76. We have ignored the transverse wave vector here as the thickness of the mesa is small and only considered the wave vector $k$ of the mode in the plane of the layer.

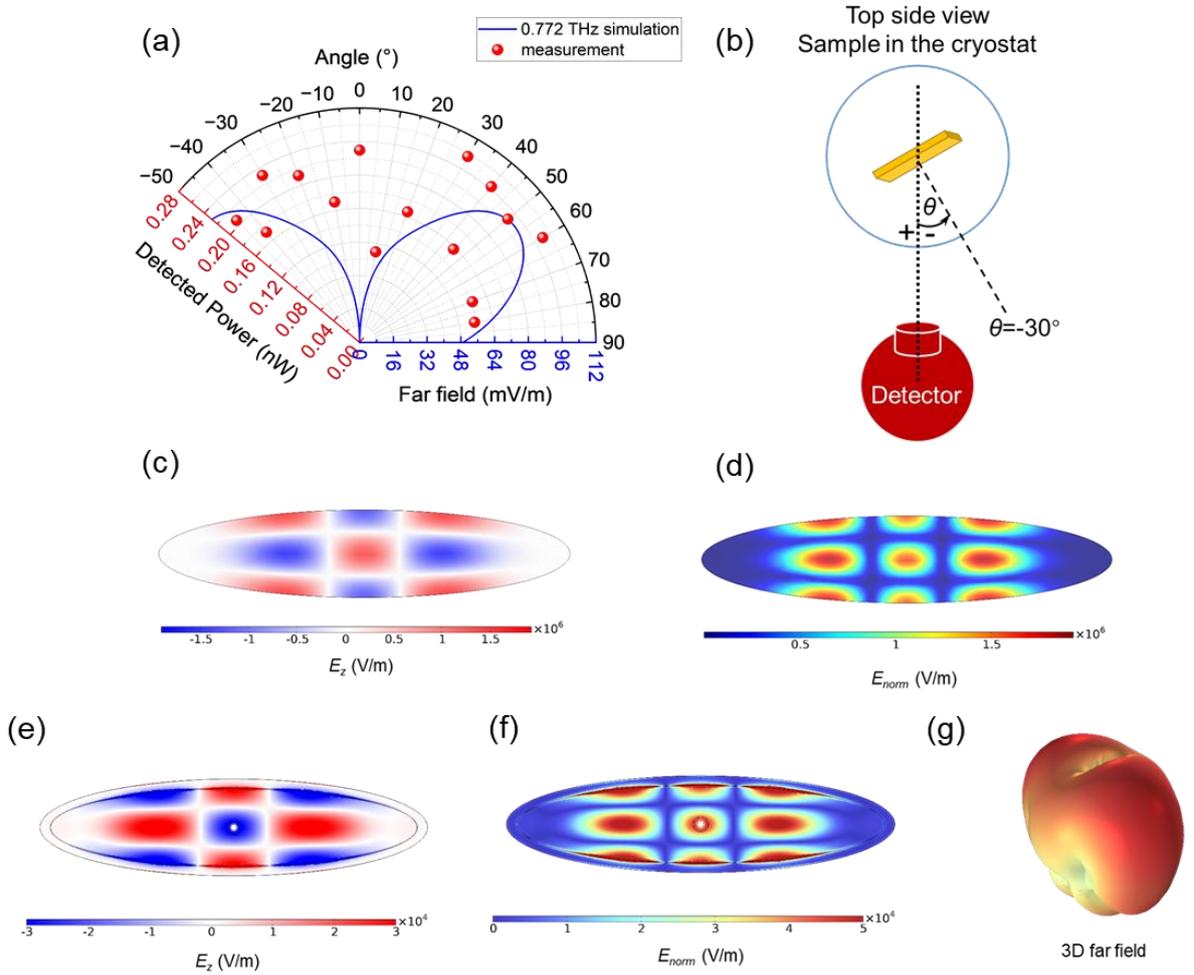

**Figure 6** (a) Angle-dependent radiation pattern of device 1 from the late 2023 measurement, showing detected power through a 10cm room temperature over-the-air link (red) at $T_b$=45 K, along with the simulated 2D far-field patterns at 0.772 THz from COMSOL with central feedline. (b) The illustration of measurement angle $\theta$, which is the angle between the normal line of the device and the bolometer detector. The power is calculated from the HEB output voltage. (c) and (d) is the $z$-direction and the normal electric field for a stand-alone elliptical antenna without feedline and groove at the eigenfrequency 798.9 GHz, respectively. (e) and (f) is the $z$-direction and the normal electric field for our device with central feedline, groove, and surrounding BSCCO at 772 GHz, respectively. (g) is the 3D far-field at the same condition as (e) and (f).

The value of $q_{mr}$ can be determined in Wolfram Mathematica by scanning the zero point of the curve for Eq.3, provided in Appendix A, Table A1. As a result, this device has the TM$_{even}$(0,1) mode at 701.66 GHz, TM$_{even}$(1,1) mode at 749.55 GHz and TM$_{even}$(2,1) mode at 798.87 GHz, which are close to the radiation frequencies. We first consider a stand-alone elliptical mesa with 1 μm thickness without a feedline, groove, or substrate. The electric field distribution at (0,1), (1,1), and (2,1) modes is provided in Fig. A10 and Fig.6 (c)-(d) by eigenfrequency simulation in COMSOL with perfectly matched frequencies at 701.45 GHz, 749.5 GHz, and 798.9 GHz.

However, the device is more complex than a single stand-alone mesa. In our simulation, we consider the remaining bottom layer (due to imperfect FIB milling) and the surrounding BSCCO crystal together with the groove cavity and the central feedline. Although the wave function and eigenfrequency for a standalone mesa suggest a (1,1) mode around 750 GHz, closely matching the device's radiation frequency, strong resonance with a similar (1,1) pattern is not observed in the device simulation within the 0.7–0.8 THz range. On the other hand, between 0.768 THz and 0.78 THz, the elliptical mesa exhibits a strong TM(2,1) mode with a matching electric field distribution and far-field pattern, as shown in Fig. 6(e)–(g) and Fig. A11. The strongest far-field pattern, plotted at 772 GHz in Fig. 6(a), closely resembles the measured radiation pattern. The observed 0.768–0.78 THz (2,1) mode resonance, which differs by approximately 19–30 GHz from the standalone mesa, is influenced by the presence of the surrounding BSCCO crystal, the groove, and the central feedline. Discrepancies between the device and the COMSOL model, including variations in the relative permittivity of BSCCO and the size and thickness of the surrounding BSCCO crystal, may contribute to the slight frequency shift between the true radiation and the simulation.

**Coherent THz photon radiation rate** Quantum applications in the THz regime, particularly THz quantum communication, require the ability to generate and precisely manipulate THz photons. Developing a coherent, stable, and tunable THz source is therefore essential, as it directly influences the feasibility of high-speed and secure quantum information transfer and processing. To analyse radiation at the photon scale, we determined the photon number $N_{photon}$ emitted by THz source over a fixed time interval using the following calculation:

$$N_{photon} = \frac{E_d}{E_p} \qquad (6)$$

with the emission energy reaching the detector

$$E_d = Pt \tag{7}$$

and the single photon energy with frequency $f$

$$E_p = hf \tag{8}$$

Here, $P$ is the power related to the HEB output voltage, as defined in Eq. (1), using the setup illustrated in Fig. A2 (b), and $t$ is time in seconds. Figures 7 (a) and (b) show the photon number detected by the HEB from device 1 over picosecond timescales at different bath temperatures at $\theta=0°$, shown as a function of bias voltage and frequency, respectively. As the bath temperature increases, the required bias voltage decreases. Notably, the highest photon number rate is observed at $T_b=30$ K, where, on average, more than 240 photons per picosecond (ps$^{-1}$) are detected after passing through a 10 cm over-the-air free-space channel. A higher bath temperature of 43 K reduces the radiation rate but maintains it above 50 ps$^{-1}$, supporting high-speed communication. Additionally, as shown in Fig.7 (b), the photon frequency shows a general decreasing trend with increasing temperature.

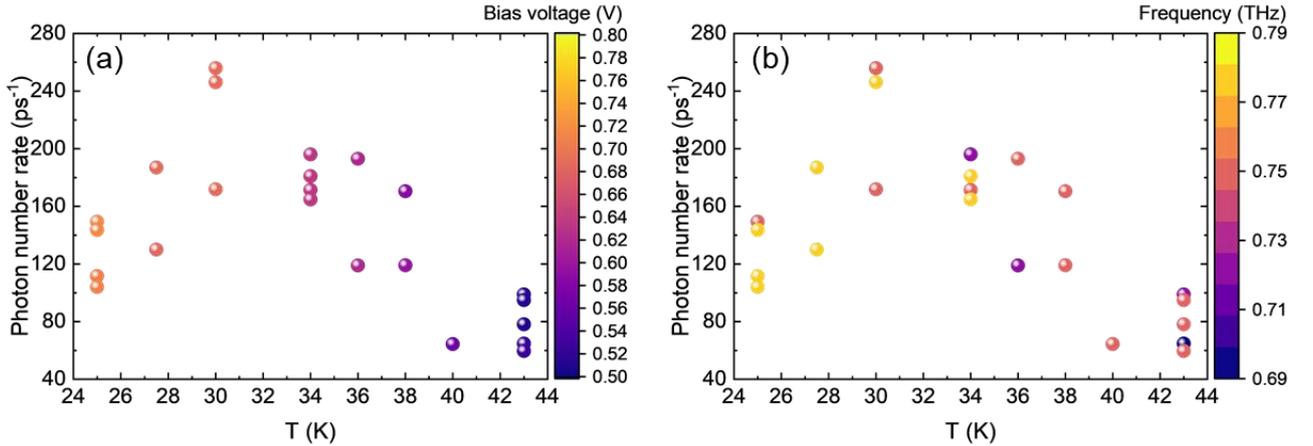

**Figure 7** The detected photon number per picosecond at different bath temperatures is shown as a function of (a) bias voltage and (b) frequency, at $\theta=0°$.

Frequency tunability at each temperature can be controlled by bias voltage or by selecting different branches in the IVCs, though this also influences the photon number, particularly in the sub-100 ps$^{-1}$ range. Additional temperature-dependent results at $\theta=30°$ and bias voltage-dependent results at $\theta=0°$ are presented in Appendix A, Fig. A7 and Fig. A8. The above discussion is based on the detected power after a ~10 cm room-temperature free-space over-the-air link. To better understand the radiation power in a vacuum environment for cryogenic platform applications, such as wireless communication within quantum computers, we calculated the radiation power and photon number near the device inside the cryostat. Considering the radiation pattern in Fig.6 (a), the average power from -45° to 80° at the detection point is ~ 0.194 nW. The measurement solid angle is 0.02 sr. Assuming an attenuation factor of 0.75 for the cryostat window made of quartz glass in HEB [28], and THz attenuation around 750 GHz is more than 1000 dB/km [50], we estimate the radiation power from the device to be ~ 250 nW inside the cryostat, with 90% transmission of the polythene window. The total radiation photon number inside the fridge is estimated to be ~503 photons per femtosecond ($fs^{-1}$), considering the radiation frequency of 750 GHz. It is noted that a radiation power of ~ 250 nW is typical for this generation of coherent THz devices, without standalone BSCCO cavities.

**Conclusion** In conclusion, we demonstrated that a chip-scale coherent THz source based on IJJs in a high-temperature superconducting BSCCO with an elliptical cavity remains operational over 11 years post-fabrication. The emission is strong enough to propagate through a nearly half-meter combined cryogenic and over-the-air channel. Changes in the doping level of the device resulted in a higher critical temperature of up to 90 K while maintaining comparable emission power to that observed over a decade ago. The thin Ag and Au layers on top of the device effectively protected against atmospheric degradation. We successfully measured the frequency spectrum with strong intensity in the 700–800 GHz range using a Michelson interferometer at room temperature and compared the radiation pattern with simulations. Additionally, we demonstrated a method for determining the transverse mode of electromagnetic radiation using the modified Mathieu function. The evaluation of THz photon generation in the 50–260 $ps^{-1}$ range highlights the device's high-speed potential for quantum applications, while also underscoring the importance of photon-level modulation and detection. While the current implementation remains at an early stage, this first estimation opens new perspectives for advancing THz quantum circuits and on-chip components based on high-temperature superconducting materials.

**Acknowledgement** This work in part was supported by the Royal Academy of Engineering (LTRF2223-19-138), Royal Society of Edinburgh International Joint Project Award, Royal Society (RGS\R2\222168), Monbukagakusho scholarship, and JST-CREST Japan. The authors thank Drs H. Minami, M. Tsujimoto, H. Asai, M. Tachiki, and T. Yamamoto for valuable discussions.

**Appendix A**

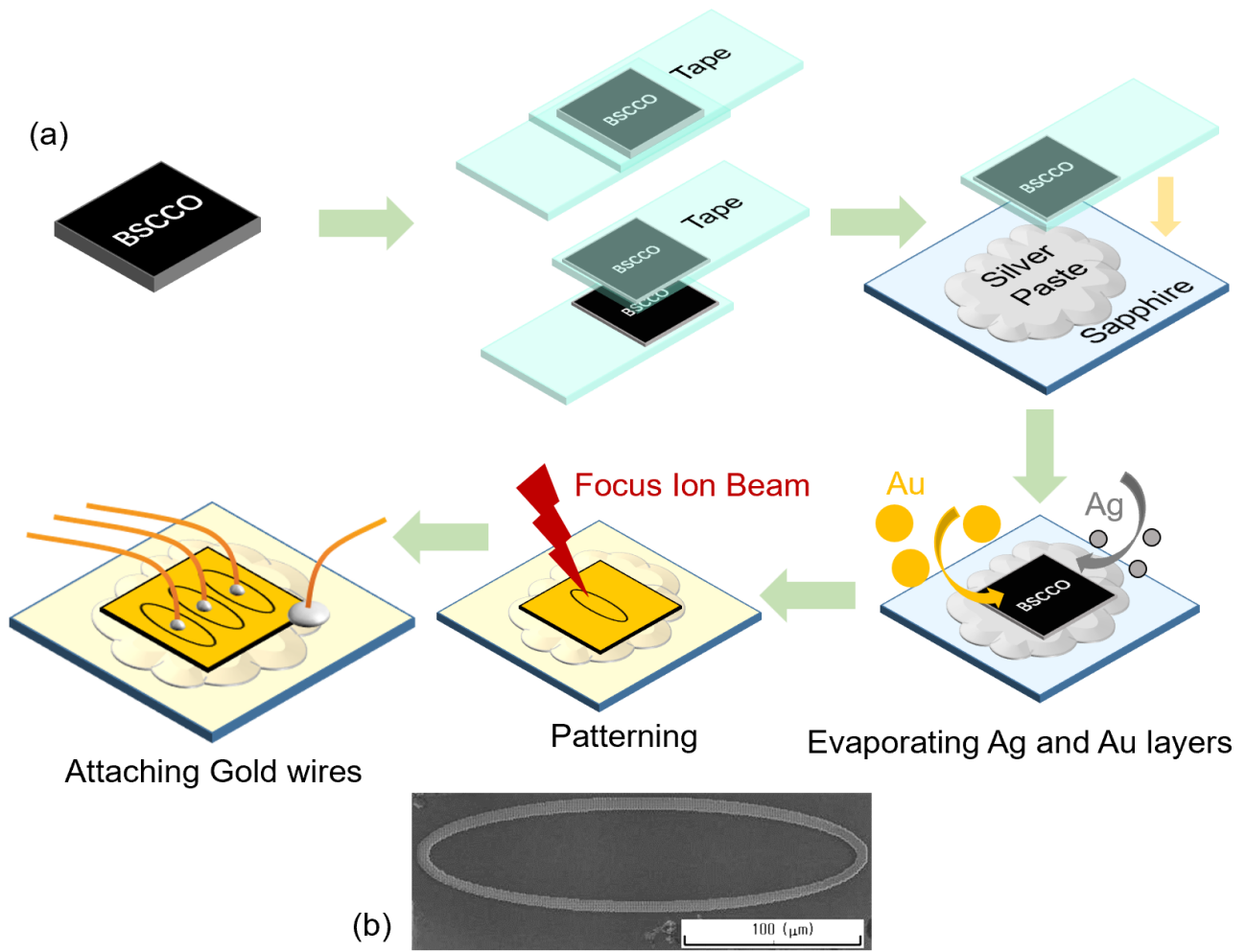

**Figure A1** (a) Schematic illustrating the fabrication process of the coherent THz device. (b) SEM image of a single coherent THz emitter with an elliptical microcavity captured in 2012.

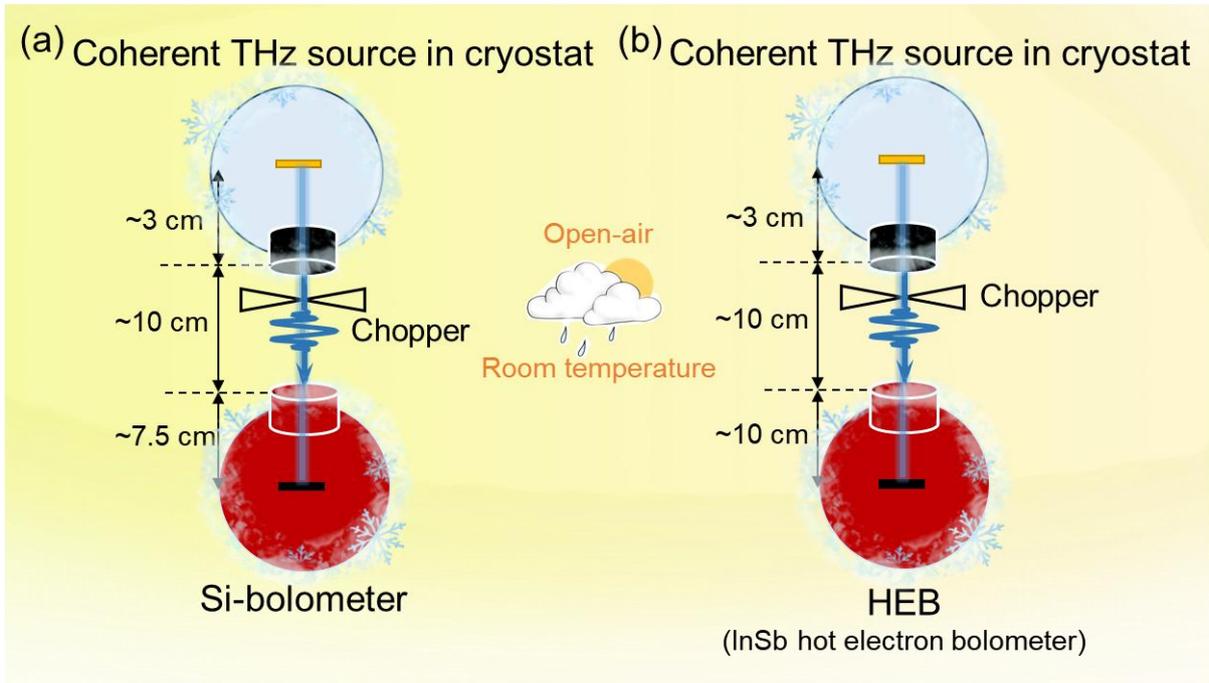

**Figure A2** Schematic diagram of the coherent THz radiation setup, including cryogenic operation, free-space transmission at room temperature, and cryogenic detection, as implemented in (a) 2012 and (b) 2023.

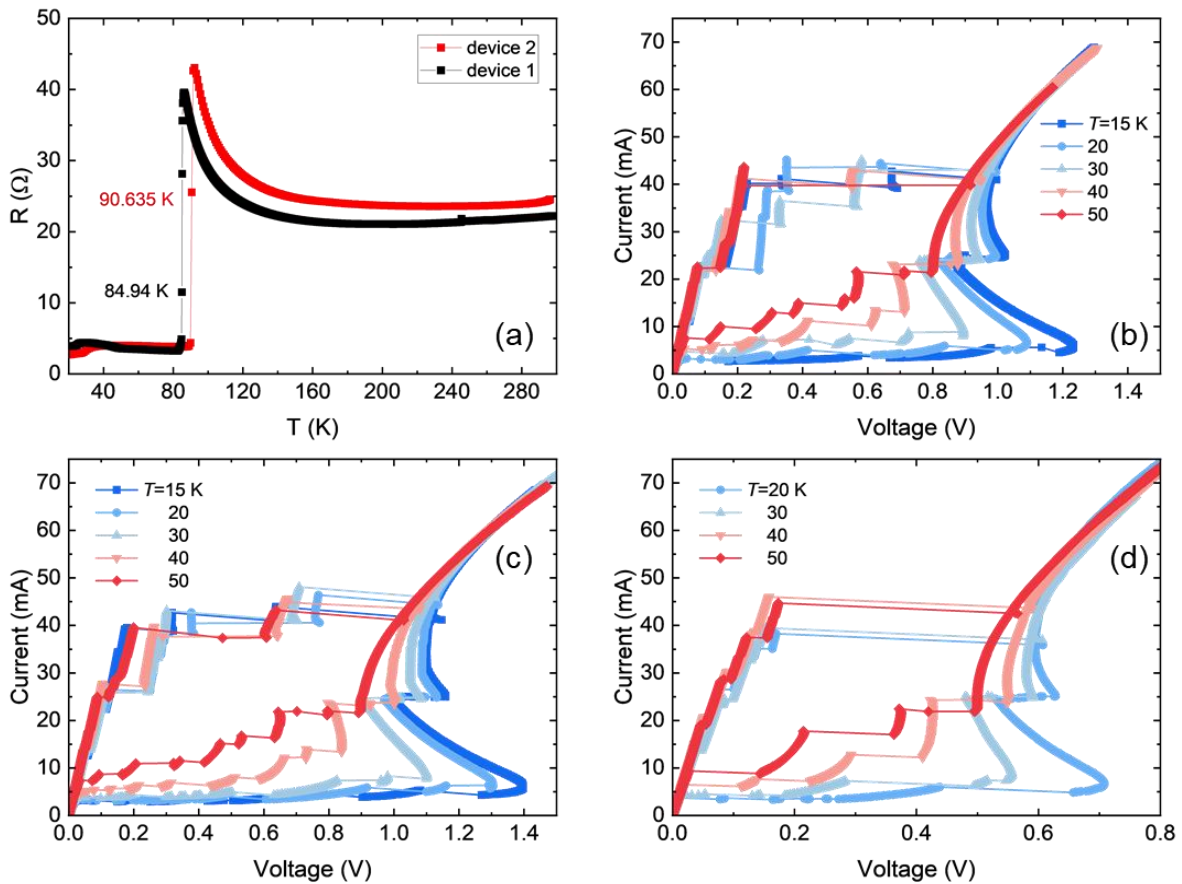

**Figure A3** (a) The temperature-dependent resistance of the elliptical cavity device 1 and device 2 in 2012. (b) The current-voltage characteristic at different bath temperatures in 2012 for device 1. (c) The current-voltage characteristic at different bath temperatures in 2012 for device 2. (d) The current-voltage characteristic at different bath temperatures in 2012 for device 3.

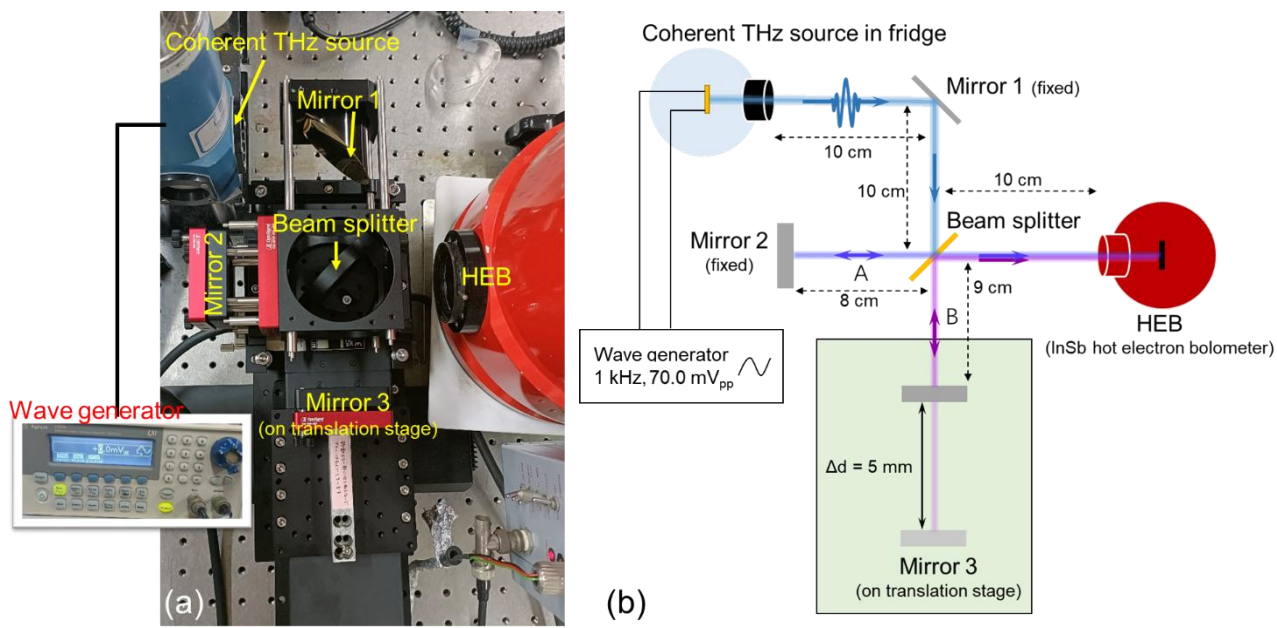

**Figure A4** (a) The photo of the THz frequency measurement setup: hybrid cryogenic-free space room temperature over-the-air THz link. (b) The schematic of the THz frequency hybrid cryogenic-free space over-the-air testbed.

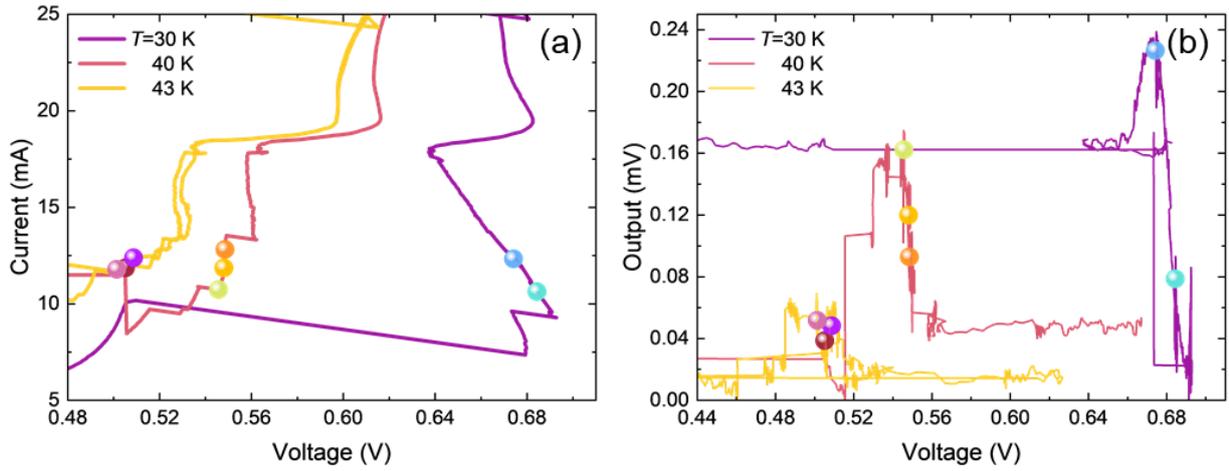

**Figure A5** (a) The current-voltage characteristic at different bath temperatures. The highlight points are bias points for frequency measurement at $\theta=30°$. (b) The output voltage from the HEB detector at different temperatures, $\theta=30°$. The highlight points are output points for frequency measurement.

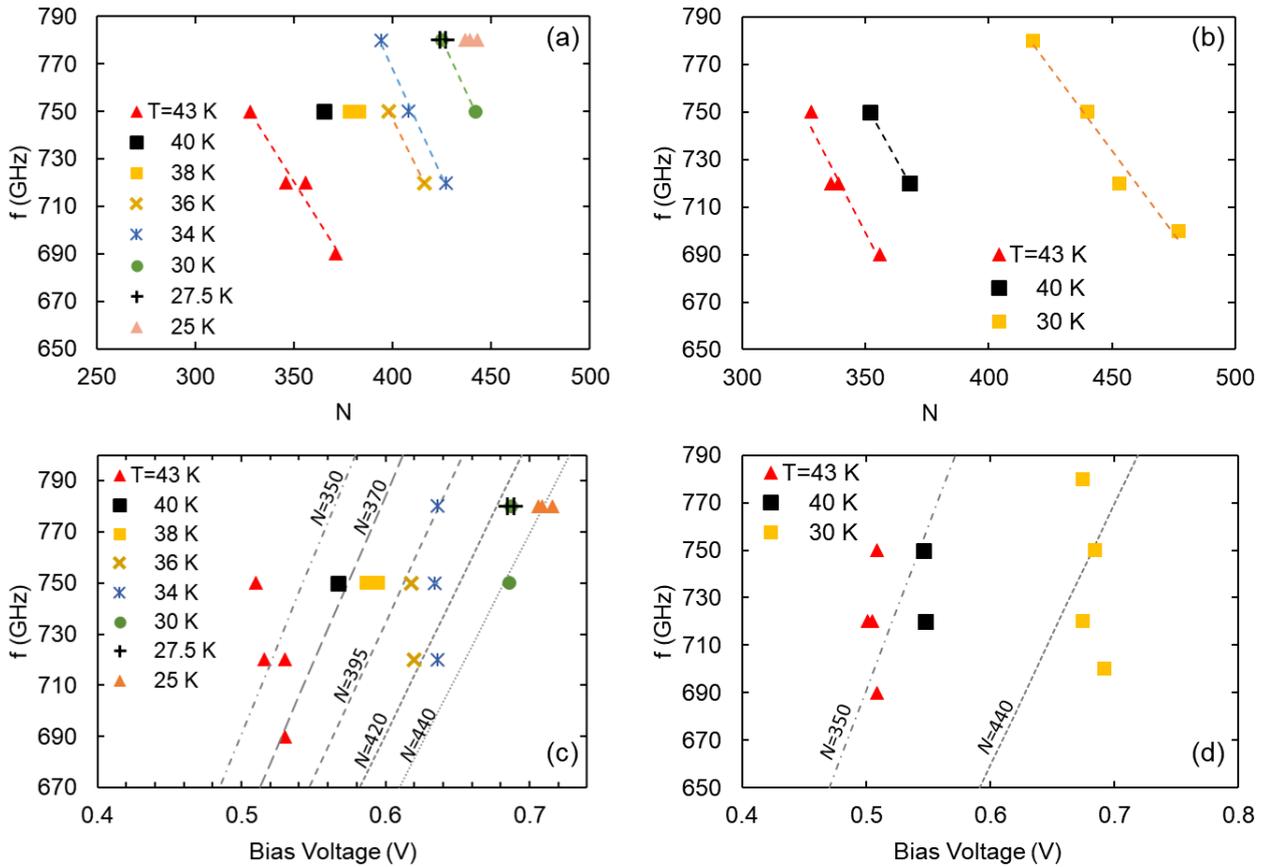

**Figure A6** The emission frequency and the number of active Josephson junctions $N$ in device 1 calculated from the ac-Josephson relation at different bath temperatures in 2023. (a) $\theta=0°$ and (b) $\theta=30°$. (c) and (d) show the frequency relations with bias voltage and the fit junction number lines at (c) $\theta=0°$ and (d) $\theta=30°$.

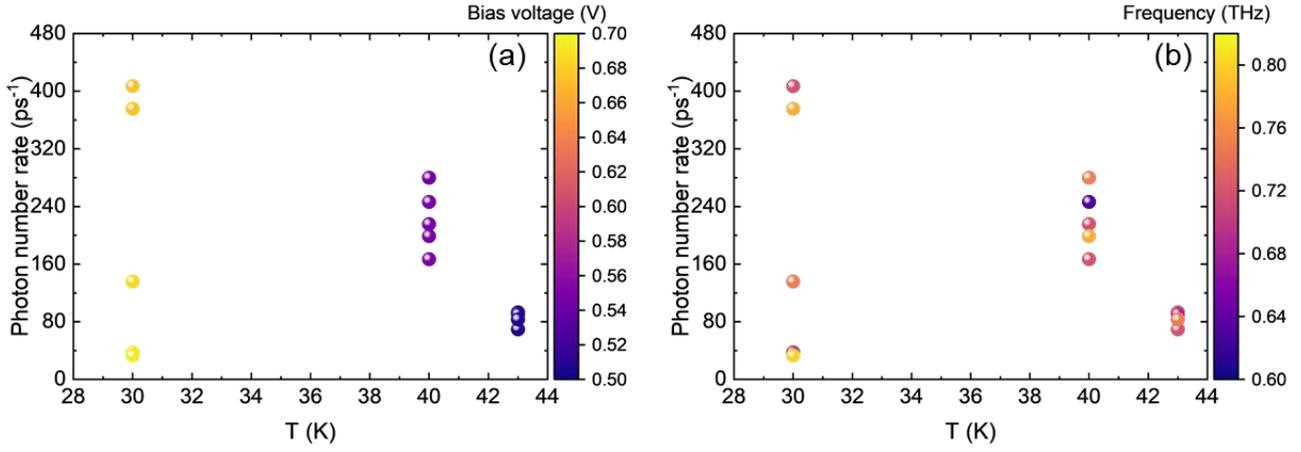

**Figure A7** The photon number detected per picosecond at different bath temperatures, with (a) varying bias voltages and (b) different frequencies, at $\theta=30°$ in 2023.

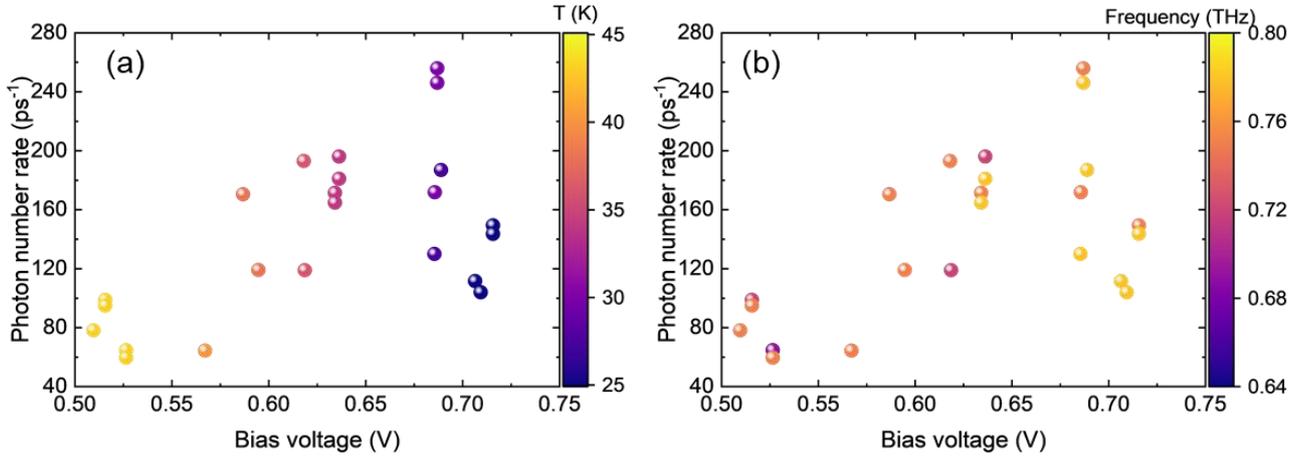

**Figure A8** The photon number approached the detector at bias voltage per picosecond, with (a) different bath temperatures and (b) different frequencies, at $\theta=0°$ in 2023.

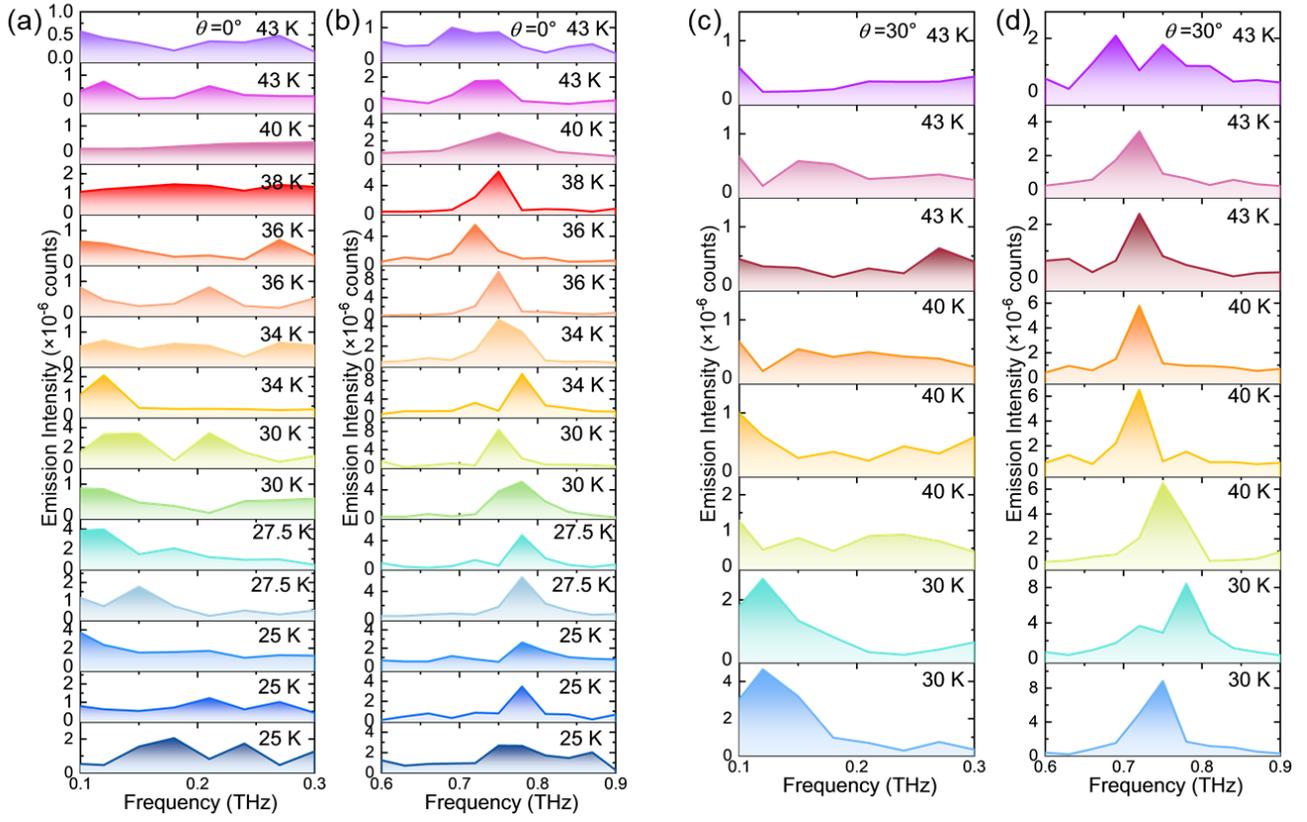

**Figure A9** Zoom-in spectra for the elliptical cavity at different bath temperatures. (a) 0.1-0.3 THz and (c) 0.6-0.9 THz, at θ=0°. The intensities are multiplied by factors of 3, 3, and 2 in 43 K and 40 K. (b) 0.1-0.3 THz and (d) 0.6-0.9 THz, at θ=30°, with intensity multipliers of 3, 2, and 3 for 43 K. The intensities in all plots are offset for clarity.

Table A1 Eigenfrequency analysis based on the modified Mathieu function.

| | *m* | *r* | *q* | *f* (GHz) |
|---|---|---|---|---|
| ∂Ce=0<br>Even solution | 0 | 1 | 55.20466 | 701.66 |
| | | 2 | 213.9617 | 1381.35 |
| | 1 | **1** | **62.99719** | **749.55** |
| | | 2 | 228.843 | 1428.59 |
| | 2 | **1** | **71.56025** | **798.87** |
| | | 2 | 244.4761 | 1476.58 |
| ∂Se=0<br>Odd solution | 1 | 1 | 14.6126 | 361 |
| | | 2 | 121.652 | 1041.59 |
| | 2 | 1 | 18.8 | 409.47 |
| | | 2 | 132.99 | 1089.05 |

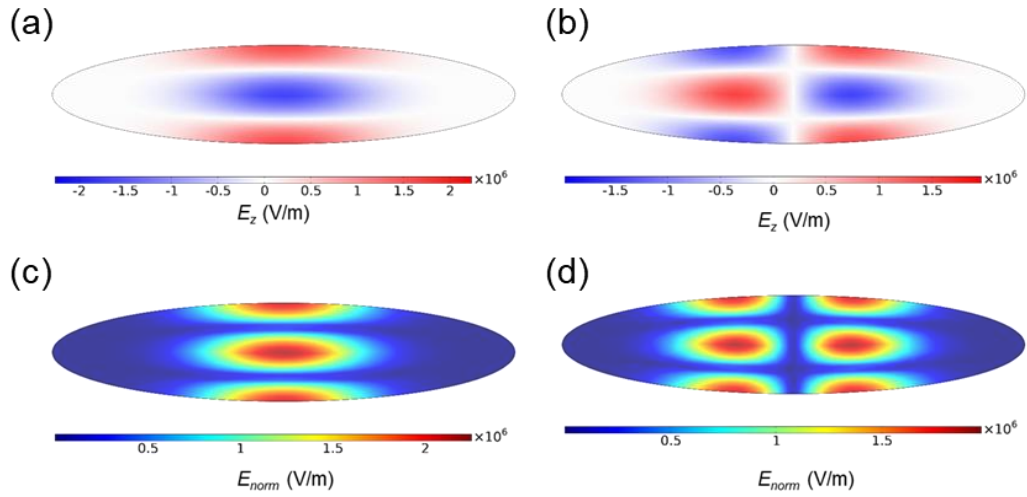

**Figure A10** (a) and (b) are the z-direction electric field distributions for a stand-alone mesa at eigenfrequencies 701.45 GHz and 749.5 GHz, respectively. (c) and (d) is the norm electric field for a stand-alone mesa at eigenfrequencies 701.45 GHz and 749.5 GHz, respectively.

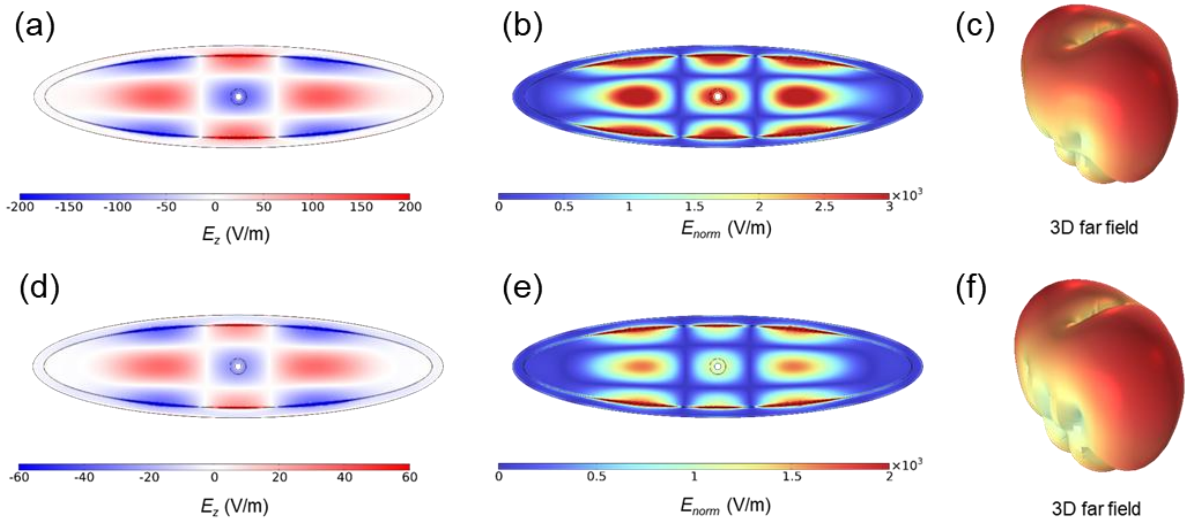

**Figure A11** (a), (b) and (c) are the z-direction electric field, norm electric field, and 3D far field for the device simulation with central feedline and groove at 768 GHz, respectively. (d), (e) and (f) are the z-direction electric field, norm electric field, and 3D far field for the device simulation with central feedline and groove at 780 GHz, respectively.

# Reference

[1] S. Shi, S. Yuan, J. Zhou, and P. Jiang, "Terahertz technology and its applications in head and neck diseases," (in eng), *iScience,* vol. 26, no. 7, p. 107060, 2023.

[2] X. Chen, H. Lindley-Hatcher, R. I. Stantchev, J. Wang, K. Li, A. Hernandez Serrano, Z. D. Taylor, E. Castro-Camus, and E. Pickwell-MacPherson, "Terahertz (THz) biophotonics technology: Instrumentation, techniques, and biomedical applications," *Chemical Physics Reviews,* vol. 3, no. 1, p. 011311, 2022.

[3] R. Wieland, O. Kizilaslan, N. Kinev, E. Dorsch, S. Guénon, Z. Song, Z. Wei, H. Wang, P. Wu, D. Koelle, V. P. Koshelets, and R. Kleiner, "Terahertz emission from mutually synchronized standalone $Bi_2Sr_2CaCu_2O_{8+x}$ intrinsic-Josephson-junction stacks," *Physical Review Applied,* vol. 22, no. 4, p. 044022, 2024.

[4] W. Nsengiyumva, S. Zhong, L. Zheng, W. Liang, B. Wang, Y. Huang, X. Chen, and Y. Shen, "Sensing and Nondestructive Testing Applications of Terahertz Spectroscopy and Imaging Systems: State-of-the-Art and State-of-the-Practice," *IEEE Transactions on Instrumentation and Measurement,* vol. 72, pp. 1-83, 2023.

[5] Z. Chen, C. Han, Y. Wu, L. Li, C. Huang, Z. Zhang, G. Wang, and W. Tong, "Terahertz Wireless Communications for 2030 and Beyond: A Cutting-Edge Frontier," *IEEE Communications Magazine,* vol. 59, no. 11, pp. 66-72, 2021.

[6] I. F. Akyildiz, C. Han, Z. Hu, S. Nie, and J. M. Jornet, "Terahertz Band Communication: An Old Problem Revisited and Research Directions for the Next Decade," *IEEE Transactions on Communications,* vol. 70, no. 6, pp. 4250-4285, 2022.

[7] C. Ottaviani, M. J. Woolley, M. Erementchouk, J. F. Federici, P. Mazumder, S. Pirandola, and C. Weedbrook, "Terahertz Quantum Cryptography," *IEEE Journal on Selected Areas in Communications,* vol. 38, no. 3, pp. 483-495, 2020.

[8] M. Zhang, S. Pirandola, and K. Delfanazari, "Millimetre-Waves to Terahertz SISO and MIMO Continuous Variable Quantum Key Distribution," *IEEE Transactions on Quantum Engineering,* pp. 1-11, 2023.

[9] N. K. Kundu, M. R. McKay, and R. K. Mallik, "Wireless quantum key distribution at terahertz frequencies: Opportunities and challenges," *IET Quantum Communication,* vol. n/a, no. n/a, 2024.

[10] M. Kutas, B. Haase, P. Bickert, F. Riexinger, D. Molter, and G. von Freymann, "Terahertz quantum sensing," *Science Advances,* vol. 6, no. 11, p. eaaz8065.

[11] T. Kashiwagi, M. Tsujimoto, T. Yamamoto, H. Minami, K. Yamaki, K. Delfanazari, K. Deguchi, N. Orita, T. Koike, R. Nakayama, T. Kitamura, M. Sawamura, S. Hagino, K. Ishida, K. Ivanovic, H. Asai, M. Tachiki, R.A Klemm, K. Kadowaki, High temperature superconductor terahertz emitters: fundamental physics and its applications, *Jpn. J. Appl. Phys.* vol. 51 p. 010113, 2012

[12] T. Kashiwagi, T. Yamamoto, H. Minami, M. Tsujimoto, R. Yoshizaki, K. Delfanazari, T. Kitamura, C. Watanabe, K. Nakade, T. Yasui, K. Asanuma, Y. Saiwai, Y. Shibano, T. Enomoto, H. Kubo, K. Sakamoto, T. Katsuragawa, B. Marković, J. Mirković, R. A. Klemm, and K. Kadowaki, "Efficient Fabrication of Intrinsic-Josephson-Junction Terahertz Oscillators with Greatly Reduced Self-Heating Effects," *Physical Review Applied,* vol. 4, no. 5, p. 054018, 2015.

[13] K. Delfanazari, H. Asai, M. Tsujimoto, T. Kashiwagi, T. Kitamura, T. Yamamoto, M. Sawamura, K. Ishida, C. Watanabe, S. Sekimoto, H. Minami, M. Tachiki, R. A. Klemm, T. Hattori, and K. Kadowaki, "Tunable terahertz emission from the intrinsic Josephson junctions in acute isosceles triangular $Bi_2Sr_2CaCu_2O_{8+\delta}$ mesas," *Optics Express,* vol. 21, no. 2, pp. 2171-2184, 2013.

[14] S. Kalhor, S. J. Kindness, R. Wallis, H. E. Beere, M. Ghanaatshoar, R. Degl'Innocenti, M. J. Kelly, S. Hofmann, H. J. Joyce, D. A. Ritchie, and K. Delfanazari, "Active Terahertz Modulator and Slow Light
21